\documentclass[aps,pra,twocolumn,superscriptaddress,nofootinbib]{revtex4-1}
\usepackage{mathrsfs}
\usepackage{amsfonts}
\usepackage{txfonts}
\usepackage{amssymb}
\usepackage{graphicx}
\usepackage{bm}
\usepackage{color}

\newcommand{\ket}[1]{|#1\rangle}
\newcommand{\bra}[1]{\langle #1|}

\begin{document}
\title{Path-shortening realizations of nonadiabatic holonomic gates}
\author{G. F. Xu}
\affiliation{Department of Physics, Shandong University, Jinan 250100, China}
\affiliation{Department of Physics and Astronomy, Uppsala University, Box 516,
Se-751 20 Uppsala, Sweden}
\author{D. M. Tong}
\email{tdm@sdu.edu.cn}
\affiliation{Department of Physics, Shandong University, Jinan 250100, China}
\author{Erik Sj\"{o}qvist}
\email{erik.sjoqvist@physics.uu.se}
\affiliation{Department of Physics and Astronomy, Uppsala University, Box 516,
Se-751 20 Uppsala, Sweden}
\date{\today}
\begin{abstract}
Nonadiabatic holonomic quantum computation uses non-Abelian geometric phases to
implement a universal set of quantum gates that are robust against control imperfections
and decoherence. Until now, a number of three-level-based schemes of
nonadiabatic holonomic computation have been put forward, and several of them
have been experimentally realized. However, all these works are based on the same class
of nonadiabatic paths, which originates from the first nonadiabatic holonomic proposal.
Here, we propose a universal set of nonadiabatic holonomic gates based on an
extended class of nonadiabatic paths. We find that nonadiabatic holonomic gates can be
realized with paths shorter than the known ones, which provides the possibility of realizing
nonadiabatic holonomic gates with less exposure to decoherence. Furthermore, inspired
by the form of this new type of paths, we find a way to eliminate decoherence from
nonadiabatic holonomic gates without resorting to redundancies.
\end{abstract}
\maketitle
\date{\today}
\section{Introduction}
In the circuit model of quantum computation \cite{nielsen01}, information is processed by
means of unitary transformations, i.e., quantum gates acting on qubits, and therefore the
central requirement of the circuit-based quantum computation is to realize high-fidelity
quantum gates. However, the practical implementation of high-fidelity quantum gates is
very challenging, mainly due to control errors and decoherence. Control errors and decoherence
are respectively induced by inaccurate manipulations and interactions with environments, and
they can lead to errors that propagate through the computation. To reduce the effects of these
error sources, different kinds of robust quantum gates have been proposed, among
which nonadiabatic holonomic gates play a prominent role.

Nonadiabatic holonomic gates are realized by using nonadiabatic non-Abelian geometric 
phases \cite{anandan88}, i.e., nonadiabatic holonomies on the Grassmann manifold 
$\mathcal{G} (N;L)$, which is the space of $L$ dimensional subspaces of an $N$ dimensional 
Hilbert space. These gates can be performed at high speed and depend only on the global 
nature of evolution paths of quantum systems, which make them robust against control errors. 
Due to these features, nonadiabatic holonomic gates have received considerable attention
and various three-level-based nonadiabatic holonomic schemes have been put forward
\cite{Sjoqvist,Abdumalikov,Feng,Arroyo,Zu,Xu3,Sjoqvist2,Herterich,
Wang,Xue2017,Li2017,Sekiguchi,Zhou2017,Hong2018,danilin18,Xu,Liang,Zhang1,Xue,Zhou,
Xue1,Zhao}. In particular, nonadiabatic holonomic gates have been demonstrated experimentally
in circuit quantum electrodynamics \cite{Abdumalikov,danilin18}, nuclear magnetic resonance 
\cite{Feng,Li2017}, and nitrogen-vacancy (NV) centers in diamond 
\cite{Arroyo,Zu,Sekiguchi,Zhou2017}.

Although impressive progress, both theoretically and experimentally, have been made
in the field of nonadiabatic holonomic quantum computation, the possibility to find nonadiabatic
paths with a larger degree of flexibility is still largely unexplored. In fact, all the above mentioned
works \cite{Sjoqvist,Abdumalikov,Feng,Arroyo,Zu,Xu3,Sjoqvist2,Herterich,
Wang,Xue2017,Li2017,Sekiguchi,Zhou2017,Hong2018,danilin18,Xu,Liang,Zhang1,Xue,
Zhou,Xue1,Zhao} are based on the same class of nonadiabatic paths, being characterized 
by an effective rotation angle of the bright state restricted to $\pi$. Thus, one interesting and
challenging topic is to find new, feasible and useful nonadiabatic paths, which is of 
significance for applications in quantum gate design. 

Here, we propose a universal set of nonadiabatic holonomic gates based on an
extended class of paths in the Grassmannian. We find that nonadiabatic holonomic gates
can be realized with paths shorter than the known ones, which provides the possibility of
realizing nonadiabatic holonomic gates with less exposure to decoherence. Furthermore,
inspired by the form of this new type of paths, we find a way to eliminate decoherence from
nonadiabatic holonomic gates without redundancies. Our proposal can be realized in 
various systems and is experimentally  feasible. To realize it, one needs only to apply pulses 
with adjustable oscillation frequency, Rabi frequency, and phase, which can be realized 
with current experimental techniques.

\section{Path-shortening scheme}
\subsection{Single-qubit gates} 
Consider a three-level system with energy eigenstates $\ket{0}$, $\ket{1}$, $\ket{e}$, and
eigenvalues $\omega_{0}$, $\omega_1$, $\omega_e$. The states $\ket{0}$ and $\ket{1}$
are used as computational states, and $\ket{e}$ is an auxiliary state. The transition between
$\ket{j}$ ($j\in\{0,1\}$) and $\ket{e}$ is induced by the laser field
${\bf E}_j(t)={\bm \epsilon}_{j}g_j(t)\cos\nu_{j}t$,
where ${\bm \epsilon}_j$ is the polarization, $g_j(t)$ is the pulse envelope
function, and $\nu_j$ is the oscillation frequency. Thus, the system Hamiltonian in
the laboratory frame is
$H_{{\rm lab}} (t)=H_s+{\bm \mu}\cdot[{\bf E}_0(t)+{\bf E}_1(t)]$,
where $H_s=-\omega_{e0}\ket{0}\bra{0}-\omega_{e1}\ket{1}\bra{1}$ is the bare Hamiltonian
with the energy of $\ket{e}$ set to zero and ${\bm \mu}$ is the electric dipole operator. We
perform a transformation to a rotating frame by using the rotation operator
$V(t)=\exp \left[-i\left(\nu_0\ket{0}\bra{0}+\nu_1\ket{1}\bra{1}\right)t\right]$, which
turns the system Hamiltonian ${H}_{{\rm lab}}(t)$ into
${H}_{{\rm rot}}(t) = \sum_{j\in\{0,1\}} [ \Omega_{ej} ( e^{i\phi_j} \ket{e}
\bra{j} + {\rm H.c.} ) - \Delta_{ej}\ket{j}\bra{j} ]$.
Here, the real-valued Rabi frequency $\Omega_{ej}$, the laser phase $\phi_j$, and
the detuning $\Delta_{ej}$ satisfy $\Omega_{ej}{e}^{i\phi_j} =
g_j(t) \bra{e}{\bm \mu\cdot{\bm \epsilon}_j}\ket{j}/2$ and
$\Delta_{ej}=\omega_{ej}-\nu_j$, respectively.
In ${H}_{{\rm rot}}(t)$, we have ignored rapidly oscillating terms (rotating wave
approximation). By assuming $\Delta_{e0}=\Delta_{e1}=\Delta$ and
making a suitable shift of zero point energy, the Hamiltonian reads
\begin{eqnarray}
{H}_{{\rm rot}} = \Delta\ket{e}\bra{e}+\Omega\big(e^{i\varphi}\ket{e}\bra{b} +
{\rm H.c.} \big),
\label{hrot}
\end{eqnarray}
where $\Omega=\sqrt{\Omega_{e0}^2 + \Omega_{e1}^2}$ and $\varphi=\phi_0$. Here,
$\ket{b}=\cos{\theta}\ket{0}+\sin{\theta}e^{i\phi}\ket{1}$ is the bright
state, with time-independent $\tan\theta={\Omega_{e1}}/{\Omega_{e0}}$ and $\phi=\phi_0-\phi_1$.
The dark state $\ket{d}=\sin{\theta}\ket{0}-\cos{\theta}e^{i\phi}\ket{1}$ is decoupled from the
dynamics.

We use Hamiltonian ${H}_{{\rm rot}}$ to realize our proposal. To see this,
we first briefly explain how nonadiabatic holonomies arise in unitary evolutions. Consider an
$N$ dimensional system containing an $L$ dimensional computational subspace
$\mathcal{S} (0)={\rm Span}\{ \ket{\phi_k(0)} \}_{k=1}^{L<N}$. The evolution operator
driven by the Hamiltonian $H(t)$ is a nonadiabatic holonomic gate acting on $\mathcal{S} (0)$
if for some time $T$ the following two conditions are satisfied:
$ {\rm (i)}  \ \sum_{k=1}^L\ket{\phi_k (T)} \bra{\phi_k (T)} =
\sum_{k=1}^L \ket{\phi_k (0)} \bra{\phi_k (0)}$ and
${\rm (ii)} \ \bra{\phi_k (t)}H(t)\ket{\phi_l (t)}=0, \ k,l = 1, \ldots ,L,$
where $\ket{\phi_k(t)} = {\bf T} \exp{[-i\int_0^tH(t')dt']} \ket{\phi_k(0)}$, with ${\bf T}$ being
time-ordering. The above two conditions  guarantee that the evolution of the computational
subspace is both cyclic (i) and geometric (ii).

To realize our proposal with Hamiltonian ${H}_{{\rm rot}}$, we first split the evolution into a
segment-pair and analyze under which requirements the system defines a nonadiabatic
holonomic gate. The segment-pair is performed on the time intervals $[0, \tau]$ and
$[\tau, T]$, where $\tau$ is the first segment ending time, coinciding with the initial time of
the second segment (we assume $\tau < T$). The Hamiltonian of the first segment reads
\begin{eqnarray}
H_1=\Delta_1\ket{e}\bra{e}+\Omega_1\big(e^{i\varphi_1}\ket{e}\bra{b_1} + {\rm H.c.} \big),
\label{h1}
\end{eqnarray}
where $\Delta_1$, $\Omega_1$, $\varphi_1$, and $\ket{b_1}$ are the detuning, the
Rabi frequency, the laser phase, and the
bright state, respectively. The dark state of $H_1$ is denoted by $\ket{d_1}$. $H_1$ is
defined in a frame associated with the rotation operator
\begin{eqnarray}
V_1(t)=\exp \left[-i\left(\nu_{1,0}\ket{0}\bra{0}+\nu_{1,1}\ket{1}\bra{1}\right)t\right],
\label{r1}
\end{eqnarray}
where $\nu_{1,0}$ and $\nu_{1,1}$ are laser frequencies satisfying $\omega_{e0}-\nu_{1,0} =
\omega_{e1}-\nu_{1,1} = \Delta_1$. The computational subspace
$\mathcal {S}(0) = {\rm Span} \{ \ket{b_1},\ket{d_1}\}$ evolves into $\mathcal {S}(\tau) =
{\rm Span} \{ U_1 \ket{b_1}, U_1 \ket{d_1}\}$ with the time evolution operator
$U_1 = e^{-iH_1 \tau}$, corresponding to the full traversal of the first segment. One finds
\begin{eqnarray}
U_1\ket{b_1}&=&e^{-i\vartheta_1\cos\eta_1}\big[(\cos\vartheta_1
+i\sin\vartheta_1\cos{\eta_1})\ket{b_1} \nonumber \\
&&-ie^{i\varphi_1}\sin\vartheta_1\sin{\eta_1}\ket{e}\big], \nonumber \\
U_1\ket{d_1}&=&\ket{d_1} ,
\label{s1}
\end{eqnarray}
where $\vartheta_1 = \sqrt{(\Delta_1/2)^2 + \Omega_1^2} \tau$ and $\tan \eta_1 =
2\Omega_1 / \Delta_1$. Since $H_1$ commutes with the corresponding time evolution operator
$e^{-iH_1t}$,
the geometric condition (ii) reduces to $\bra{j}H_1\ket{j^\prime}=0$,
with $j, j^\prime \in 0,1$. Thus, the first segment evolution
$\mathcal{S} (0) \rightarrow \mathcal{S} (\tau)$ is purely geometric as long as the
corresponding Hamiltonian has the form of ${H}_{{\rm rot}}$ in Eq.~(\ref{hrot}).

For the second segment, the Hamiltonian can be written as
\begin{eqnarray}
H_2=\Delta_2\ket{e}\bra{e}+\Omega_2\big(e^{i\varphi_2}\ket{e}\bra{b_2} + {\rm H.c.} \big),
\label{h2}
\end{eqnarray}
where as above $\Delta_2$, $\Omega_2$, $\varphi_2$, and $\ket{b_2}$ represent the detuning,
the common Rabi frequency, the laser phase, and the bright state, respectively. The dark state
of $H_2$ is $\ket{d_2}$. $H_2$ is defined in a frame associated with the rotation operator
\begin{eqnarray}
V_2(t)=\exp \left[-i\left(\nu_{2,0}\ket{0}\bra{0}+\nu_{2,1}\ket{1}\bra{1}\right)t\right],
\label{r2}
\end{eqnarray}
where $\nu_{2,0}$ and $\nu_{2,1}$ are the laser frequencies, satisfying the relation
$\omega_{e0} -\nu_{2,0}=\omega_{e1}-\nu_{2,1}=\Delta_2$. Since we allow for
$\Delta_2 \neq \Delta_1$,
the rotating frames of the two segments, i.e., the rotation operators $V_1 (\tau)$ and
$V_2 (\tau)$, can be different. In order to compensate for this difference,
the initial computational subspace of the second segment should be taken as
$V_2(\tau)V_1^\dag(\tau)\mathcal {S}(\tau)$. By using Eqs.~(\ref{r1}),
(\ref{s1}), and (\ref{r2}), the basis states of $V_2(\tau)V_1^\dag(\tau)\mathcal {S}(\tau)$
read
\begin{eqnarray}
\ket{\psi_1} & = & e^{i[(\Delta_2-\Delta_1)\tau-\vartheta_1\cos\eta_1]}
\big[(\cos\vartheta_1 + i \sin\vartheta_1\cos{\eta_1})\ket{b_1}
\nonumber \\
 & & -ie^{i[\varphi_1+(\Delta_1-\Delta_2)\tau]}\sin\vartheta_1\sin{\eta_1}\ket{e}\big],
\nonumber \\
\ket{\psi_2} & = & e^{i(\Delta_2-\Delta_1)\tau}\ket{d_1}.
\label{s2}
\end{eqnarray}
In the following, we first show how to make sure the geometric condition (ii) is satisfied
for the initial subspace ${\rm Span} \left\{ \ket{\psi_1},\ket{\psi_2} \right\}$ of the second
segment, followed by a demonstration of the cyclic condition (i) for the segment pair.

Since $H_2$ and $e^{-iH_2t}$ commute, the geometric condition $\textrm{(ii)}$ turns into
the constraints $\bra{\delta}H_2\ket{\delta^\prime}=0$, where $\ket{\delta},\ket{\delta^\prime}
\in \{\ket{\psi_1},\ket{\psi_2}\}$. We discuss these four constraints one by one.
First we note that, since $\ket{d_1} \in {\rm Span} \{ \ket{b_1},\ket{d_1}\} = {\rm Span}
\{ \ket{b_2},\ket{d_2}\}$, we can rewrite the basis states $\ket{\psi_1}$ and $\ket{\psi_2}$
in Eq.~(\ref{s2}) as
\begin{eqnarray}
\ket{\psi_1} & = & c_1\ket{e}+c_2\ket{b_2}+c_3\ket{d_2},
\nonumber \\
\ket{\psi_2} & = & c_4\ket{b_2}+c_5\ket{d_2},
\label{s3}
\end{eqnarray}
where $c_j$ are complex numbers. By combining Eqs.~(\ref{h2}) and (\ref{s3}), we find
\begin{eqnarray}
\bra{\psi_1}H_2\ket{\psi_2}=c_1^{\ast}c_4\Omega_2{e^{i\varphi_2}}.
\end{eqnarray}
The coefficient $c_1$ must be nonzero in order for the evolution along the first segment
to be noncyclic. Thus, the only nontrivial solution of $\bra{\psi_1} H_2 \ket{\psi_2}=0$ is
$c_4=0$. This implies that $\ket{d_1}$ and $\ket{d_2}$ are the same up to a global phase.
Thus, $\bra{\psi_1}H_2\ket{\psi_2}=0$ requires that $H_2$ takes the form
\begin{eqnarray}
H_2=\Delta_2\ket{e}\bra{e}+\Omega_2\big(e^{i\varphi_2}\ket{e}\bra{b_1} + {\rm H.c.} \big) ,
\label{h3}
\end{eqnarray}
which in turn implies that the geometric constraints $\bra{\psi_2}H_2\ket{\psi_1}=0$ and
$\bra{\psi_2}H_2\ket{\psi_2}=0$ are satisfied.

It remains to check the constraint $\bra{\psi_1}H_2\ket{\psi_1}=0$.
To this end, we define $U_2 = e^{-iH_2 (T-\tau)}$, being the time evolution operator
corresponding to the full traversal of the second segment, and temporarily assume 
that the evolution satisfies condition (i), i.e., the computational subspace performs
cyclic evolution on $[0,T]$. By using Eqs.~(\ref{s2}) and (\ref{h3}), we first find that
$U_2 \ket{\psi_2} = e^{i\varphi_{d_1}} \ket{d_1}$ for some global phase $\varphi_{d_1}$.
Thus, the cyclic condition (i) entails that
\begin{eqnarray}
U_2\ket{\psi_1}=e^{i\varphi_{b_1}}\ket{b_1},
\label{cycle}
\end{eqnarray}
where $\varphi_{b_1}$ is a global phase. By combining $[H_2,U_2]=0$ and Eq.~(\ref{cycle}),
one finds
\begin{eqnarray}
\bra{\psi_1}H_2\ket{\psi_1} = \bra{\psi_1}U_2^\dag{H_2}U_2\ket{\psi_1}
=\bra{b_1}H_2\ket{b_1} = 0 . \label{condition}
\end{eqnarray}
In other words, $\bra{\psi_1} H_2 \ket{\psi_1}=0$ is satisfied as long as the evolution
generated by $H_2$ satisfies the cyclic condition (i).

To complete the analysis, we now demonstrate how to satisfy the cyclic condition (i). To
this end, we rewrite Eq.~(\ref{cycle}) as $U_2^\dag\ket{b_1} = e^{-i\varphi_{b_1}} \ket{\psi_1}$,
and evaluate the left-hand side
\begin{eqnarray}
U_2^\dag\ket{b_1} & = & e^{i\vartheta_2\cos\eta_2}\big[(\cos\vartheta_2
-i\sin\vartheta_2\cos{\eta_2})\ket{b_1}
\nonumber \\
 & & + ie^{i\varphi_2}\sin\vartheta_2\sin{\eta_2}\ket{e}\big],
 \label{i}
\end{eqnarray}
where $\vartheta_2 = \sqrt{(\Delta_2/2)^2+(\Omega_2)^2} (T-\tau)$ and $\tan\eta_2 =
2\Omega_2 / \Delta_2$. It is noteworthy that the parameters $\vartheta_2$, $\eta_2$, and
$\varphi_2$ can be chosen independently of the corresponding parameters of the first pulse.
By combining Eqs.~(\ref{s2}) and (\ref{i}) with $U_2^\dag\ket{b_1} = e^{-i\varphi_{b_1}}
\ket{\psi_1}$, one finds that as long as
\begin{eqnarray}
\left| \sin\vartheta_2 \sin{\eta_2} \right| = \left| \sin\vartheta_1\sin{\eta_1} \right|,
\label{cycle1}
\end{eqnarray}
the populations satisfy $|\bra{e}U_2^\dag\ket{b_1}|^2=|\langle{e}\ket{\psi_1}|^2$ and
$|\bra{b_1}U_2^\dag\ket{b_1}|^2=|\langle{b_1}\ket{\psi_1}|^2$. Thus, under the condition
in Eq.~(\ref{cycle1}), it only remains to adjust the phase $\varphi_2$ in order for the evolution
to satisfy the cyclic condition (i). Since both $U_2^\dag\ket{b_1}$ and $\ket{\psi_1}$
contain $\ket{e}$, it follows that $\left| \sin\vartheta_1\sin{\eta_1} \right|$ and
$\left| \sin \vartheta_2 \sin{\eta_2} \right|$ are nonzero. If furthermore
$\left| \sin \vartheta_2\sin{\eta_2} \right| = \left| \sin\vartheta_1\sin{\eta_1} \right| <1$,
it follows that $\varphi_2$ is a relative phase between the states $\ket{e}$ and $\ket{b_1}$,
and should be chosen as
\begin{eqnarray}
\varphi_2 & = & \varphi_1+(\Delta_1-\Delta_2)\tau-a-\sum_{j=1}^2\arg(\cos\vartheta_j
\nonumber \\
 & & + i\sin\vartheta_j\cos{\eta_j}),
\label{varphi1}
\end{eqnarray}
where $a=\pi$ if $\sin\vartheta_2\sin{\eta_2}=\sin\vartheta_1\sin{\eta_1}$, and $a=0$
if $\sin\vartheta_2\sin{\eta_2}=-\sin\vartheta_1\sin{\eta_1}$. If $\left| \sin\vartheta_2 \sin{\eta_2}
\right| = \left| \sin\vartheta_1\sin{\eta_1} \right| = 1$, the phase $\varphi_2$ becomes a
global phase and there is no special constraint on its value.

Until now, we have found the requirements under which the system defines a
nonadiabatic holonomic gate. The holonomic gate reads
\begin{eqnarray}
U=\exp(i\beta)\ket{b_1}\bra{b_1}+\ket{d_1}\bra{d_1},
\end{eqnarray}
where $\beta=\sum_{j=1}^2\arg(\cos\vartheta_j+i\sin\vartheta_j\cos\eta_j)-\vartheta_j
\cos\eta_j$ if $\left| \sin \vartheta_j\sin{\eta_j} \right| <1$, and $\beta=\varphi_1-\varphi_2 +a $
if $\left| \sin \vartheta_j\sin{\eta_j} \right| =1$. The corresponding nonadiabatic path
reads
\begin{eqnarray}
\mathcal {S}_{\{\ket{b_1},\ket{d_1}\}}\rightarrow\mathcal {S}_{\{\ket{\psi_1},\ket{\psi_2}\}}
\rightarrow \mathcal {S}_{\{e^{i\beta}\ket{b_1},\ket{d_1}\}},
\end{eqnarray}
where $\mathcal {S}_{\{\ket{A},\ket{B}\}}$ denotes the subspace spanned by states $\ket{A}$
and $\ket{B}$. In fact, by using the derived conditions, we can realize more flexible nonadiabatic
paths. According to Eq.~(\ref{condition}), ${H}_2$ can be used to drive the system for time
$\tau \leq t < T$, while keeping the holonomic feature. In this case, the corresponding
evolution is ${S}_{\{\ket{\psi_1},\ket{\psi_2}\}}\rightarrow{S}_{\{\ket{\psi_1^\prime},
\ket{\psi_2^\prime}\}}$, where $\ket{\psi_1^\prime}$ is a superposition state of $\ket{e}$ and
$\ket{b_1}$, and $\ket{\psi_2^\prime}=e^{i\xi}\ket{d_1}$ with $\xi$ being the compensation
phase. Then a Hamiltonian $H_3$ having the form of Eq.~(\ref{hrot}) can be constructed to
continue this path. By repeating, we can realize nonadiabatic paths of the form 
\begin{eqnarray}
\mathcal{S}_{\{\ket{b_1},\ket{d_1}\}} & \rightarrow & \mathcal{S}_{\{\ket{\psi_1},\ket{\psi_2}\}}
\rightarrow \mathcal{S}_{\{\ket{\psi_1^\prime},\ket{\psi_2^\prime}\}} \rightarrow \cdots
\nonumber \\
 & \rightarrow & \mathcal{S}_{\{e^{i\gamma}\ket{b_1},\ket{d_1}\}}, \label{path}
\end{eqnarray}
where $e^{i\gamma}$ is a geometric phase factor.

Following  Eq.~(\ref{path}), one can construct nonadiabatic paths different from
the previous ones. This can be seen from the evolution of
the bright state $\ket{b}$. Since $\ket{b}$ evolves in the subspace ${\rm Span}\{\ket{b}, \ket{e}\}$,
its evolution can be viewed by the effective Bloch sphere $\mathcal {B}$ with $\ket{b}$
and $\ket{e}$ being its poles. For the previous schemes 
\cite{Sjoqvist,Abdumalikov,Feng,Arroyo,Zu,Xu3,Sjoqvist2,Herterich,Wang,Xue2017,Li2017,
Sekiguchi,Zhou2017,Hong2018,danilin18,Xu,Liang,Zhang1,Xue,Zhou,Xue1,Zhao}, the 
bright state $\ket{b}$ evolves along
a path on $\mathcal {B}$ that corresponds to an effective total rotation angle
$\sum_j \vartheta_j = \sum_j \sqrt{(\Delta_j/2)^2 + \Omega_j^2} T_j  = \pi$, $T_j$
being the duration time of the $j$th segment. On the
other hand, for the paths in Eq.~(\ref{path}), the effective rotation angle of $\ket{b}$ can be
smaller than $\pi$. We illustrate this with an example. Consider a path containing two
segments, where the first segment is induced by
\begin{eqnarray}
{H}_1=\Delta_1\ket{e}\bra{e} + \Omega_1\big(e^{i\varphi_1}\ket{e}\bra{b}+e^{-i\varphi_1}
\ket{b}\bra{e}\big) .
\end{eqnarray}
We assume the effective rotation angle $\vartheta_1 = \sqrt{(\Delta_1/2)^2 + \Omega_1^2}
T_1=\pi/3$, and the ratio of $2\Omega_1/\Delta_1$ is $3/4$. According to the derived conditions,
the Hamiltonian that induces the second segment, given ${H}_1$, can be chosen to have the form
\begin{eqnarray}
{H}_2 = \Omega_2 \big( e^{i\varphi_2} \ket{e} \bra{b} + e^{-i\varphi_2} \ket{b} \bra{e} \big)
\end{eqnarray}
with the phase $\varphi_2=\varphi_1-0.6\pi-\arg(5+3\sqrt{3}i)$ and the effective rotation angle
$\vartheta_2 = |\Omega_2| T_2 \approx 0.24\pi$. The realized nonadiabatic holonomic
gate for this path is $\mathscr{U} = e^{i\alpha} \ket{b}\bra{b} + \ket{d}\bra{d}$ with
$\alpha\approx\pi/18$. According to the above calculations, the total effective rotation
angle $\vartheta_1 + \vartheta_2$ for the bright state $\ket{b}$ is thus about $0.57 \pi$,
which is significantly smaller than $\pi$. More importantly, the above example
shows that shorter paths can be realized, which provides the possibility of realizing 
nonadiabatic holonomic gates with less exposure to unwanted decoherence effects.

\subsection{Dynamical decoupling}
Much effort has been paid focusing on realizing nonadiabatic holonomic gates with
less exposure to decoherence \cite{Xu,Liang,Zhang1,Xue,Zhou,Zhao,xu14,sun16,zhang18}. These
works resort to redundancies which consist of encoding logical qubits with sets of
physical qubits. 
Here, inspired by the multi-segment form of Eq.~(\ref{path}), we propose a new approach to 
eliminate decoherence from nonadiabatic holonomic gates without using redundancies. Since 
the paths in Eq.~(\ref{path}) contain many segments, we consider interleaving these segments
with a dynamical decoupling sequence \cite{Zhang2014}. Consider a three-level system
experiencing decay, induced by the system-environment interaction Hamiltonian
\begin{eqnarray}
H_{SE} = (\ket{e} \bra{0} + \ket{0} \bra{e})\otimes{E_{e0}} +
(\ket{e} \bra{1} + \ket{1} \bra{e})\otimes{E_{e1}},
\end{eqnarray}
where $E_{e0}$ and $E_{e1}$ are environment operators. For $H_{SE}$, the decoupling
group can be taken as $\{I, g_1, g_2, g_3\}$, where $I$ is the identity operator,
$g_1 = \ket{1}\bra{1}-\ket{e}\bra{e}-\ket{0}\bra{0}$, $g_2=\ket{e}\bra{e}-\ket{0}\bra{0}-\ket{1}\bra{1}$,
and $g_3=\ket{0}\bra{0}-\ket{e}\bra{e}-\ket{1}\bra{1}$. The dynamical decoupling sequence
is   $I\rightarrow{g_1}\rightarrow{g_3}\rightarrow{g_1}\rightarrow{g_3}$, with $g_1 =
e^{-i\pi(\ket{e}\bra{0}+{\rm H.c.})}$ and $g_3=e^{-i\pi(\ket{e}\bra{1}+{\rm H.c.})}$. Consider
a four-segment path $\mathcal {S}_{\{\ket{b_1},\ket{d_1}\}}\rightarrow
\mathcal {S}_{\{\ket{\psi_1},\ket{\psi_2}\}}\rightarrow \mathcal {S}_{\{\ket{\psi_1^\prime},
\ket{\psi_2^\prime}\}}
\rightarrow\mathcal {S}_{\{\ket{\psi_1^{\prime\prime}},\ket{\psi_2^{\prime\prime}}\}}\rightarrow
\mathcal {S}_{\{e^{i\gamma}\ket{b_1},\ket{d_1}\}}$, with the segment Hamiltonians being
${H}_1$, ${H}_2$, ${H}_3$, and ${H}_4$. After the interleaf, the whole evolution reads
\begin{eqnarray}
I\rightarrow{{H}_1^\prime}\rightarrow{g_1}\rightarrow{{H}_2^\prime}\rightarrow{g_3}
\rightarrow{{H}_3^\prime}\rightarrow{g_1}\rightarrow{{H}_4^\prime}\rightarrow{g_3}. \label{sequence}
\end{eqnarray}
One can see that if $H_1^\prime=H_1$, ${H}_2^\prime=g_1{H}_2g_1$, ${H}_3^\prime =
g_2{H}_3g_2$, and ${H}_4^\prime=g_3{H}_4g_3$, then the evolution in Eq.~(\ref{sequence})
is equivalent to  ${H}_1\rightarrow{H}_2\rightarrow{H}_3\rightarrow{H}_4$. One can verify
that $H_1^\prime$, ${H}_2^\prime$, ${H}_3^\prime$, and ${H}_4^\prime$ still have the form
of Eq.~(\ref{hrot}). More importantly, the evolution driven by them remains holonomic. For
example, at the beginning of the evolution generated by ${H}_2^\prime$, the computational 
subspace is $g_1\mathcal {S}_{\{\ket{\psi_1},\ket{\psi_2}\}}$. Thus, 
$\mathcal {S}_{\{\bra{\psi_1},\bra{\psi_2}\}}
g_1{H}_2^\prime{g_1 \mathcal {S}_{\{\ket{\psi_1},\ket{\psi_2}\}}} =
\mathcal{S}_{\{\bra{\psi_1}, \bra{\psi_2}\}}{H}_2{\mathcal{S}_{\{\ket{\psi_1},\ket{\psi_2}\}}}=0$
and the holonomic feature is preserved. One can similarly use dynamical decoupling to
reduce dephasing of our holonomic scheme. In this case, the interaction Hamiltonian reads
\begin{eqnarray}
H_{SE}^\prime =\ket{0}\bra{0}\otimes{E_{0}}
+\ket{1}\bra{1} \otimes{E_{1} + \ket{e}\bra{e}\otimes{E_{e}}},
\end{eqnarray}
were $E_{0}$, $E_{1}$, and $E_{e}$ are environment operators. One can use the pulse 
$P=e^{-i\frac{\pi}{\sqrt{2}}[\ket{e}(\bra{0}-\bra{1})+{\rm H.c.}]}$ to create a common environment  
that protects the computational subspace. $P$ preserves the form of the Hamiltonian in 
Eq.~(\ref{hrot}) and the holonomic feature. Thus, the dynamical decoupling idea can be 
used to eliminate dephasing. It is noteworthy that the used decoupling pulses for $g_1$, 
$g_3$, and $P$ also have the geometric feature, which makes the whole decoherence 
eliminating method geometric.

\subsection{Two-qubit gates}
We next show that our paths can be used to realize two-qubit nonadiabatic holonomic
gates too. We use the NV center electron spin as the target qubit and one nearby $^{13}$C 
nuclear spin as the control qubit. Both the electron and nuclear spin are
polarized through optical pumping, which can be confirmed by optically detected 
magnetic resonance spectroscopy. The spins are interacting with each other through 
hyperfine and dipole couplings. By applying state-selective microwave pulses, one 
can couple the electronic spin-triplet ground states $\ket{0}$, $\ket{1}$, $\ket{a}$ 
conditionalized on the nuclear spin states $\ket{\uparrow}$, $\ket{\downarrow}$. With 
microwave fields with adjustable oscillation frequency, Rabi frequency, and phase, one 
can realize the Hamiltonian
\begin{eqnarray}
H_{j} & = & \ket{j}\bra{j} \otimes \left[ \Delta_j \ket{a} \bra{a} \right.
\nonumber \\ 
 & & \left. + 
\left( \Omega_{0,j} \ket{a} \bra{0} + \Omega_{1,j} \ket{a} \bra{1} + {\rm H.c.} \right) \right] 
\end{eqnarray}
with $j=\uparrow$ or $\downarrow$, detunings $\Delta_{j}$, and $\Omega_{0,j},\Omega_{1,j}$ 
complex-valued Rabi frequencies \cite{Zu,Liangzt,Duyx}. By alternating $H_{\uparrow}$ 
and $H_{\downarrow}$ so as to generate a nonadiabatic multi-segment path, one can realize 
the two-qubit nonadiabatic holonomic gate
\begin{eqnarray}
U_{ne} = \sum_{j=\uparrow,\downarrow} \ket{j} \bra{j} \otimes U_j , 
\end{eqnarray}
where $U_j$ are unitary holonomic operators acting on the states $\ket{0}$ and $\ket{1}$. 
$U_{ne}$ may entangle the nuclear and electronic spin qubits ${\rm Span} \{ \ket{\uparrow},
\ket{\downarrow} \}$ and ${\rm Span} \{ \ket{0},\ket{1} \}$, respectively, if $U_{\uparrow}$ and 
$U_{\downarrow}$ are different. It is noteworthy that the above two-qubit gates can also be 
protected by our decoherence eliminating method.

\section{Discussion}
To realize our proposal, Eq.~(\ref{varphi1}) is a central condition, which guarantees that 
subsequent segments match each other. When it is satisfied, the bright state evolves 
cyclically and acquires a purely geometric phase factor that translates into a non-Abelian 
holonomy via the dependence of the bright state on the laser parameters $\theta,\phi$ 
\cite{Sjoqvist}. Similarly, when realizing geometric gates in two-level systems, the basis 
states evolve cyclically and each acquires an Abelian geometric phase. Here, multi-segment 
paths, e.g., orange-slice-shaped paths, can be used (see, e.g., Ref. \cite{thomas11}), for 
which conditions similar to Eq.~(\ref{varphi1}) exist to make sure the segments match. 
Considering the feasibility of Abelian geometric gates, the central condition 
Eq.~(\ref{varphi1}) is expected to be experimentally feasible.

In actual experiments, phases and detunings can typically be implemented with high accuracy,
while imperfect control of duration or strength of pulses is hardest to deal with \cite{low14}. 
Generally, imperfect control errors are proportional to the effective rotation angle $\sum_j 
\vartheta_j$ unless there exists cancelation between the segment errors. Clearly, for
non-split paths, such cancelation is not possible because there is only one segment. For
non-split paths, the effective rotation angle is always $\pi$, while for ours, it can be less
than $\pi$. This implies that for split paths, even in the cases where the segment errors
do not cancel, the errors may have a less effect because the total
effective rotation angle is smaller. In addition, we have also shown how to interleave the
multi-segment paths with dynamical decoupling sequences. This provides the possibility
to reduce decoherence without using any additional qubits. Thus, our proposal can be used
to increase the fidelity of nonadiabatic holonomic gates.

In our calculations, we assume the detuning is time-independent and therefore the used
pulses need to be square pulses in order to preserve the purely geometric feature of the 
evolution. In fact, this assumption can be relaxed and we illustrate this with the NV center 
and the $^{87}$Rb cold atom. For the NV center, the states $\ket{0}$, $\ket{1}$ and $\ket{e}$ 
are mapped into the Zeeman components $\ket{m=-1}$, $\ket{m=+1}$ and $\ket{m=0}$, 
respectively. The transitions from $\ket{m=-1}$ and $\ket{m=+1}$ to $\ket{m=0}$ are coupled 
by a microwave field whose frequency, amplitude and phase are adjusted by mixing with 
an arbitrary-waveform generator. As a result, the Hamiltonian in Eq.~(\ref{hrot}) with the 
detuning being time-dependent can be realized \cite{Zu,Liangzt}. For the $^{87}$Rb cold 
atom, $\ket{0}$, $\ket{1}$ and $\ket{e}$ can be mapped into the Zeeman sublevels of 
$\ket{F=1, m_F=-1}$, $\ket{F=1, m_F=1}$ and $\ket{F=1, m_F=0}$, respectively. Zeeman 
sublevels with a quantum number difference $\Delta m_F = \pm1$ are coupled by radio 
frequencies or the two-photon Raman transition ($\sigma^+-\pi$). A second-order Zeeman 
effect is applied to introduce an inhomogeneous splitting between the sublevels, which 
allows individual control of both levels. In this case, the detuning can also be time-dependent 
\cite{Duyx,Lin,Lin2009,Lin2011}. The described implementations require a high degree 
of pulse control, but should be within reach with current experimental technologies.

\section{Conclusions}
In conclusion, we have proposed a universal set of nonadiabatic holonomic gates based on
an extended class of nonadiabatic paths. Specifically, we show how to realize an extended 
class of nonadiabatic multi-segment paths and develop nonadiabatic holonomic gates based 
on them. We find that these gates can be realized with paths shorter than 
the known ones, which provides the possibility of realizing nonadiabatic holonomic gates 
with less exposure to decoherence. Furthermore, inspired by the multi-segment form of this 
new type of paths, we find a way to eliminate decoherence from nonadiabatic holonomic 
gates without redundancies. Our proposal can be realized in various systems and are feasible 
in experiment.

\section*{Acknowledgments}
G.F.X. acknowledges support from the National Natural Science Foundation of China
through Grant No. 11605104, from the Future Project for Young Scholars of Shandong
University through Grant No. 2016WLJH21, and from the Carl Tryggers Stiftelse (CTS)
through Grant No. 14:441. D.M.T. acknowledges support from the National Natural Science
Foundation of China through Grant No. 11575101, and from the National Basic Research
Program of China through Grant No. 2015CB921004. E.S. acknowledges financial support
from the Swedish Research Council (VR) through Grant No. 2017-03832.

\end{document}